\documentclass[pra,twocolumn,10pt,aps,nofootinbib,superscriptaddress,showkeys,showpacs,longbibliography]{revtex4-1}
\usepackage{amsmath,amsfonts,amssymb,amstext,amscd,amsthm,dsfont}
\usepackage{color}
\usepackage{braket,tikz}
\usetikzlibrary{arrows}
\usepackage{soul,xcolor}
\usepackage{gensymb}
\usepackage[normalem]{ulem}

\usepackage[T1]{fontenc}
\usepackage{epsfig} 
\usepackage{amsmath} 
\usepackage{pifont}
\usepackage{graphicx}
\usepackage{color}
\usepackage{placeins}
\usepackage{float}
\usepackage{soul,xcolor}
\usepackage[export]{adjustbox}
\usepackage{multirow}
\usepackage[utf8]{inputenc}
\usepackage{tabularx,colortbl}
\usepackage{bbold}
\usepackage{braket}
\usepackage{titlesec}
\usepackage{placeins}
\usepackage{epstopdf}
\usepackage{upgreek}

\usepackage{graphicx}

\usepackage{braket}
\usepackage{placeins}

\begin{document}

\title{Current Trends in Quantum Optics}

\author{Subhashish Banerjee}
\email{subhashish@iitj.ac.in}
\affiliation{Indian Institute of Technology Jodhpur, India.}

\author{Arun Jayannavar}
\email{jayan@iop.res.in}
\affiliation{Institute of Physics, Bhubaneswar, India.}


\begin{abstract}
\large{Here we review some of the recent developments in Quantum Optics. After a brief introduction to the historical development of the subject, we discuss some of the modern aspects of quantum optics including atom field interactions, quantum state engineering, metamaterials and plasmonics, optomechanical systems,  PT (Parity-Time) symmetry in quantum optics as well as  quasi-probability distributions and quantum state tomography. Further, the recent developments in topological  photonics is briefly discussed. The potent role of the subject in the development of our understanding of quantum physics and modern technologies is brought out.}
\end{abstract}
\maketitle 
\large

\section{Introduction}  

Light is intimately connected to existence of all forms of life. The systematic study of light is known as optics and could be traced historically to \cite{AlHaytham}. The notion of light as an electromagnetic field was made clear by Maxwell resulting in his celebrated work on what is now known as Maxwell equations. Basically, he showed that electromagnetic fields in vacuum propagate at the speed of light.
The next step in the history of this subject were the questions raised by the Michealson-Moreley experiment and the Rayleigh-Jeans catastrophe associated with black-body radiation. The former lead to the development of special theory of relativity and the later to Plancks resolution, which provided the first seeds for the field of quantum mechanics.\par
	 The notion of photon, essentially considering it as a particle, was first realized in Einstein's work on the photo-electric effect.
	 A large number of phenomena, related to optics,  could be explained by invoking the concept of photons involving a classical electromagnetic field along with vacuum fluctuations. It was soon realized, in the wake of experimental developments, that in order to understand the full potential of the photon, one needs to treat it quantum mechanically. This lead to the development of the quantum theory of radiation, which, in turn, was the precursor to quantum optics \cite{GerryKnight}.
	 The photon provides an example of a simple quantum state labelled by, say, its hoizontal and vertical polarizations as
	                              \begin{equation*}
	                              \ket{\psi} = c_1 \ket{H} + c_2 \ket{V}.
	                              \end{equation*}
	       Here, $\ket{\psi}$, pronounced as \textit{ket psi}, denotes the state of the system and is a vector  and is an example of a \textit{superposition} state. Also, $c_1$ and $c_2$ are in general, complex numbers such that $|c_1^2| + |c_2|^2 =1$. The states $\ket{H}$ and $\ket{V}$ can exist simultaneously with probabilities $|c_1|^2$ and $|c_2|^2$, respectively.    This is the \textit{surprising}  content of quantum mechanics that makes it stand apart from classical physics. This feature can be used as a resource for achieving things not possible in the classical realms. Further, it should be noted that in quantum mechanics, we deal with operators characterizing the observables. For example, energy is represented by the operator $\hat{H}$ called the Hamiltonian of the system.  Experiments typically involve making measurements of the operator on the state vector and what is obtained as a result of the measurement is the eigenvalue of the operator. A simple example of measurement would be a projection operator that acts on the state of the system to be measured and take it to the final state.  A basic tenet of quantum mechanics is that the results of the process of measurement are probabilistic in nature, and is known as the Born rule.\par
	         Quantum measurements are described, in general, by a collection $\{M_m\}$ of  measurement operators, acting on the state space of the system being measured. The index $m$ refers to the measurement outcomes that may occur in the experiment. If the state of the quantum system is $\ket{\psi_i}$ immediately before the measurement, then the probability that result $m$ occurs is $\langle \psi_i | M_m^\dagger M_m | \psi_i \rangle$. Here, $\langle \psi_i | M_m^\dagger$ is the hermitian conjugate of the vector $M_m | \psi_i \rangle$. The measurement leaves the system in the state 
	           \begin{equation}
	           \ket{\psi_i} \rightarrow \ket{\psi_f} = \frac{M_m \ket{\psi_i}}{\sqrt{\langle \psi_i | M_m^\dagger M_m | \psi_i \rangle}}.
	           \end{equation}
           The measurement operators satisfy the completeness condition $\sum_{m}^{} M_m^\dagger M_m = \mathbb{1}$, expressing the fact that probabilities sum up to one. For those operators $M_m$ where $M_m^\dagger = M_m$ and $M_m^2 = M_m$, this reduces to the projection operators mentioned above.
	
	 Quantum optics provides tools to study foundations of quantum mechanics with precision and is the cause of a number of developments in quantum information and quantum technology.
	  An important development in quantum optics came with the formulation of coherent states \cite{glaub,sudar}, which basically ask the question of what are the states of the field that most nearly describes a classical electromagnetic field. This was originally introduced by Schrodinger \cite{Schrodinger1926}.  Classically an electromagnetic field, light, has a well defined amplitude and phase, a picture that changes in the quantum mechanical scenario. A field in a coherent state is a minimum-uncertainity state with equal uncertainities in the two variables, in this context, often termed as the two quadrature components.
	 This was followed by the development of squeezed states, where fluctuations in one quadrature component are reduced below that of the corresponding coherent state.\par
	  In the quest  for characterizing the nonclassical behavior of photons, anti-bunching and sub-Poissonian statistics were investigated. The coherent field is represented by Poissonian photon statistics, in which the photons tend to distribute themselves uniformly, i.e., the variance of the photon distribution is equal to its mean. In contrast to this, in sub-Poissonian statistics,  the variance is less than the mean and was, first experimentally demonstrated in  \cite{HongOuMandel}. The deviation from Poissonian statistics is quantified by the Mandel $Q_M$ parameter, discussed below in the context of quantum correlations. A closely related albeit distinct phenomenon is that of photon antibunching. Photon bunching is the tendency of photons to distribute themselves in bunches, such that when light falls on a photo detector, more photon pairs are detected close together in time than further apart. In contrast to this, in antibunching, the probability of photon pairs detection further apart is greater than that for pairs close together. Both the phenomena of sub-Poissonian statistics and photon antibunching are purely quantum in nature.\par
	  A major technological development due to quantum optics has been the development of LASER (Light Amplification by Stimulated Emission of Radiation), where coherent states play an essential role.
	  Quantum entanglement, which is quintessence of the  quantum world, has been experimentally demonstrated with photons and has proved to be a potential resource  in quantum computation  and communication. Spontaneous Parametric Down Conversion (SPDC) is a frequently used process in quantum optics which is very useful for generating squeezed states, single and entangled photons.

Having sketched the roots of quantum optics, we will, in the remaining part of this article, try to provide a flavor for some of the modern developments in the field. In this context, the following topics will be discussed: (a) Atom-Field interactions, (b) Quantum state engineering, (c) Metamaterials and Plasmonics, (d)  Optomechanical systems, (e)   PT (Parity-Time) symmetry in quantum optics, (f)  Quasi-probability distributions and quantum state tomography and (g) topological photonics.


\section{Atom-Field Interactions}
A cannonical model of quantum optics is the Rabi model which deals with the response of an atom to an applied field, in resonance with the atom's natural frequency.  Rabi considered the problem of a spin-half magnetic dipole undergoing precessions in a magnetic field. He  obtained the probability of  a spin-half atom  flipping from $\ket{0} = (1~0)^T$ or $\ket{1} = (0~1)^T$ to the states $\ket{1}$ or $\ket{0}$, respectively, by an applied radio-frequency magnetic field.  Here $T$ stands for the transpose of the row matrix. In the context of quantum optics, Rabi oscillations would imply oscillations of the atom, considered as a two-level system, between the upper and lower levels under the influence of the electromagnetic field. The Rabi model lead to the development of the well-known Jaynes Cumming model, which is a generalization of the semi-classical  Rabi model  in the sense that it invokes the interaction of the atom and the single mode of the  electromagnetic field in a cavity. In contrast to the semiclassical Rabi model, here the electromagnetic field is quantized.\par The Jaynes-Cummings model  has provided a platform for numerous investigations, including those related to quantum computation. However, one needs to keep in mind that the quantum mechanical coherences are subject to decay due to interactions with the surroundings \cite{SBbook}. This leads to losses such as decoherence, i.e., loss of coherence and dissipation which implies loss of energy. Hence, it is imperative that one has a good understanding of these processes. Experimental progress in this direction is discussed next.
\subsection{Experimental studies on decoherence models}

In a series of   beautiful Ion Trap experiments, the  Wineland group \cite{turchette} induced decoherence, the loss of coherence due to interaction with the surroundings, and decay by coupling the atom to engineered reservoirs in which the coupling to, and the state of the environment were controlled.  Here, the basic tool used was the coherent control of atoms’ internal states to deterministically prepare superposition states and extend this control to the external (motional) states of atoms. The Haroche group \cite{haroche} used  Cavity Quantum Electrodynamics (QED), i.e., the controlled interaction of atom with electromagnetic field in an appropriate cavity to bring out various aspects of decoherence. Microwave photons trapped in a superconducting cavity constitute an ideal system to realize some 	of the thought experiments that form the foundations of quantum physics. The interaction of 	these trapped photons with Rydberg atoms, effectively two-level atoms, crossing the cavity illustrates fundamental aspects of measurement theory.

\begin{figure*}[ht!]
	\includegraphics[width=120mm]{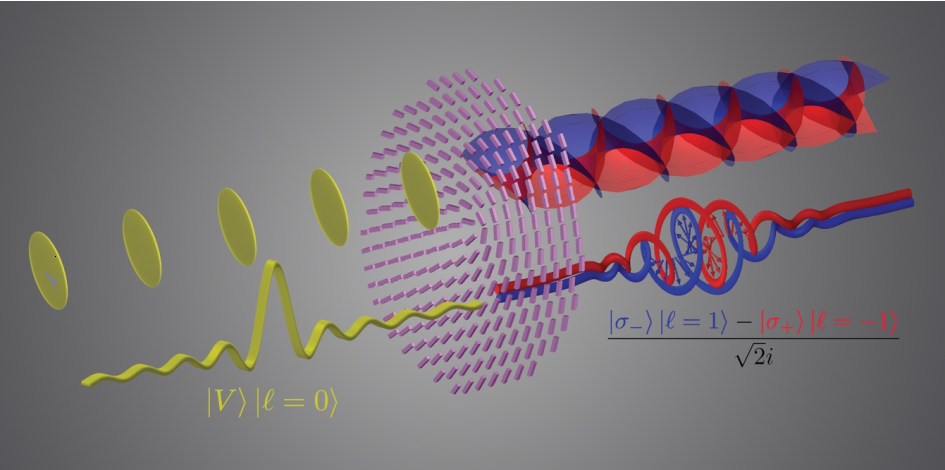}
	\caption{A single photon prepared in vertical linear polarization arrives from left, as illustrated by the yellow electric field amplitude. This photon carries zero orbital angular
		momentum, as illustrated by the yellow flat phase fronts. The single photon passes through the metasurface comprising dielectric nano-antennae (purple), and exits as a
		quantum entangled state, depicted as a superposition of the red and blue electric field amplitudes and with the corresponding vortex phase fronts opposite to one another \cite{MetaEnt}.}
	\label{metamat1}
\end{figure*}

\begin{figure}[ht!]
	\includegraphics[width=80mm]{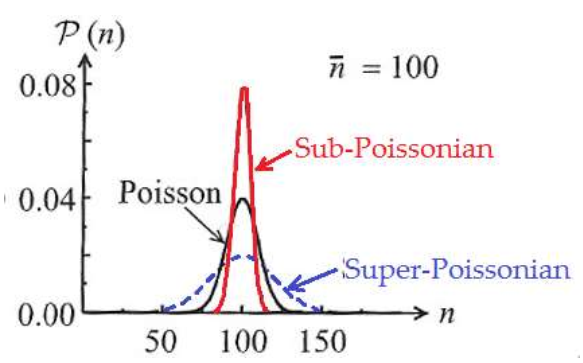}
	\caption{Probability distribution}
	\label{Poisson}
\end{figure}

\subsection{Quantum Correlations}

Here we discuss the generation of various facets of quantum correlations, including entanglement, arising from atom-field interactions. The widely studied among quantum correlations is the non separability of the quantum state of a system comprised of two or more subsystems. In the simplest scenario, consider two systems with state vectors $\ket{\psi}$ and $\ket{\phi}$. A simple choice for the combined state of the combined system would $\ket{\psi} \otimes \ket{\phi}$ where $\otimes$ denotes the tensor product. However, one can think of  states of the form $(\ket{\psi} \otimes \ket{\phi} \pm \ket{\phi} \otimes \ket{\psi})/\sqrt{2}$. By no means can one write this state as a product of two arbitrary states. Such states are called  entangled states, a well known example of which would be the Bell states \cite{nc}. A simple illustration generating a photon entangled state is shown in Fig. \ref{metamat1}. For two optical modes $\hat{a}$ and $\hat{b}$, sufficient condition for entanglement is given by the Hillery-Zubairy criteria: (i) $\langle \hat{a}^\dagger \hat{a} \hat{b}^\dagger \hat{b} \rangle < | \langle \hat{a} \hat{b}^\dagger \rangle|^2$ (ii) $\langle \hat{a}^\dagger \hat{a} \rangle \langle \hat{b}^\dagger \hat{b} \rangle < |\langle \hat{a} \hat{b} \rangle|^2$.
Another important nonclassical correlation exhibited  by light is the sub-Poissonian statistics, see Fig. \ref{Poisson}. The coherent state of light is the closest classical description in which  the probability of finding $n$ photons has Poissonian distribution: $\mathcal{P}_n = \frac{\bar{n}^n}{n!} e^{- \bar{n}}.$
Here $\bar{n}$ is the average photon number and is equal to the variance, which in this case is $ \Delta n^2 = \bar{n} $. Fluctuations are expected to increase the departure from the mean value and consequently one would have what is called super Poissonian light for which $ \Delta n^2 > \bar{n} $, an example is the thermal light. However, it is found that light under certain circumstance shows sub-Poissonian behavior, i.e., $ \Delta n^2 < \bar{n} $. This phenomenon has no classical analog and is a quantum effect. Mandel parameter defined by 
$Q_M = (\Delta n^2 - \bar{n}) /\bar{n}$ measures the departure from the Poissonian statistics. Therefore, we have $Q_M < 0$, $Q_M = 0$ and $Q_M > 0$ for sub-Poissonian, Poissonian and super-Poissonian light, respectively. 


\section{Quantum State Engineering}

Quantum state engineering (QSE) comprises of three processes: (1)
preparation, (2) detection and (3) reconstruction of quantum states. In the recent
past, QSE has appeared as the epithet for various proposals and experiments
preparing interesting states of the quantized electromagnetic field and
atomic systems. Their motivation is the potential application of nonclassical states to tasks such as, teleportation, quantum computation and communication, quantum cryptography and quantum lithography, among others. These harness the power of quantum
correlations which occupy a central position in the quest for understanding
and harvesting the utility of quantum mechanics. A field state having holes in
its photon number distribution (PND) corresponds to a nonclassical state. A
challenge in QSE is to make such holes by controlling their
position and depths in the PND. A class of states in quantum optics,
of relevance to the  present context, are the intermediate states such as the
binomial states (BS), formed by an interpolation of the number state, characterized by the number of photons in the state, with the
coherent state, as discussed above. The
concept of hole burning has been extended to the atomic domain as well.
Here the generation of spin squeezing proceeds via hole burning of selected
Dicke states, counterparts of the photonic number states, out of an atomic
coherent state, counterparts of the usual coherent states, prepared for a
collection of $N$ two$-$level atoms or ions. The atoms or ions of the atomic
coherent state are not entangled, but the removal of one or more Dicke
states generates entanglement, and spin squeezing occurs for some ranges
of the relevant parameters. Spin squeezing in a collection of two-level atoms
or ions is of importance for precision spectroscopy.
\begin{figure*}[ht!]
	(a)	\includegraphics[width=50mm]{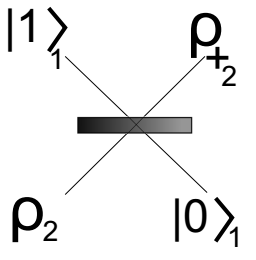}
	(b)	\includegraphics[width=50mm]{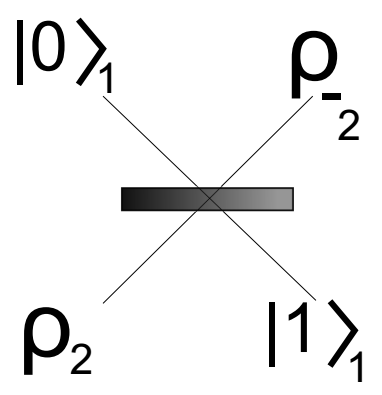}
	\caption{Beam splitter model for (a) photon addition in which a photon is added to the state $\rho_2$ and (b) photon subtraction ends up taking one photon from the state $\rho_2$, as a result of beam splitter interaction.}
	\label{BS}
\end{figure*}

QSE can be studied using linear optics; in the from of photon added or subtracted  using beam splitters (BSs), as illustrated in Fig \ref{BS}. An isomorphism can be set up between BS operations and angular
momentum relations. Squeezed states are examples of nonclassical multiphoton states of light
and can be generated from coherent states. Of special emphasis is the engineering of
nonclassical states of photons and atoms, such as, Fock states, macroscopic
superposition states, multiphoton generalized coherent and squeezed states, that are relevant for applications of forefront aspects of modern quantum mechanics. In recent times, advances have been made by combining linear and nonlinear
optical devices allowing realization of multiphoton entangled states of the
electromagnetic field, either in discrete or in continuous variables, that are relevant for
applications to efficient quantum information processing. Multiphoton quantum states are
carriers of information, and the manipulation of these by the methods of quantum
optics, has brought about a strong interface between the fields of quantum information and
quantum optics.


\section{Metamaterials and Plasmonics}
Metamaterials are structured composite materials with periodic subwavelength sized unit cells.  The subwavelength nature of the unit cells does not permit the  wave to resolve individual unit cells and permits the description of the composite material as an effective medium described by effective medium properties such as permittivity and permeability.
Metamaterials are  artificially fabricated materials, usually comprising of  nanoscale structures designed to respond to light in different ways. They have been heavily studied in last two decades and have been used to demonstrate a wealth of fascinating phenomena ranging from negative refractive index to super resolution imaging. In recent times, metamaterials have emerged as a new platform in quantum optics and have been used to carry out a number of  important experiments using single photons. Sophisticated macro-scale structures have been used to create photon entanglement \cite{Kwait}. Recent  advancements in on-chip quantum photonic circuits have lead to the development of integrated entangled photon sources. The \textit{metasurfaces} made of high refractive index dielectric are resistive to plasmonic decoherence and loss. Silicon based \textit{metasurfaces} with nearly 100$\%$ efficiency makes them candidates for quantum optics and quantum information applications. A simple illustration of a dielectric metasurface generating entanglement between the spin and the orbital angular momentum of photons is shown in Fig. \ref{metamat1}.\par 
 Plasmonics is a rapidly developing field at the boundary of physical optics and condensed matter physics. It studies phenomena induced by and associated with surface plasmons: elementary polar excitations bound to surfaces and interfaces of good nanostructured metals.\par 
  Electronic systems governed by the rules of quantum mechanics are hard to understand principally due to the strong Coloumb interactions between electrons. This makes many interesting systems difficult to comprehend via simple models as the many-body interactions and effects cannot be  ignored nor properly described in a simplistic manner.  The Maxwell
 	 equations that govern electromagnetism and the Schrodinger equation of  quantum mechanics reduce to the same Helmholtz equations  under  certain circumstances. This can be used to great effect to use light  as a test-bed for quantum effects as the  photon-photon interactions can be very small and can be  accurately controlled  via non-linear interactions of a medium. Electromagnetics has greatly benefitted in past two decades from developments in metamaterials and plasmonics. Metamaterials having resonant unit cells have highly dispersive effective medium parameters whereby novel properties and phenomena not usually found in nature become accessible. The most famous examples are negative
 	 refractive index, perfect lenses with sub-diffraction image resolution, perfect absorbers of light and media with extremely large anisotropy resulting in even hyperbolic dispersion for light. Similarly, plasmonics that concerns with the study and manipulation of surface electromagnetic waves on the surfaces of metals (electronic plasmas) and metamaterials,  gives rise to enormous possibilities on structured surfaces. With great control that becomes possible on the dispersion of the two-dimensional surface plasmons as well as their
 	 coupling to radiative modes, plasmonics offers immense promise for miniaturization of optical devices that are projected to supplant electronic devices due to the enormous  bandwidths at optical frequencies.\par 
 	  It is known that surface plasmon waves preserve the entanglement of twin photons, which is understood by the coherent nature of the surface plasmon waves inspite of the coupling to dissipative modes in the plasmonic medium.  Plasmonic systems are becoming more important for quantum information purposes. Particularly as the interaction volumes become smaller and subwavelength, the interaction of the surface plasmons with emitting molecules or quantum dots becomes increasingly governed by quantum mechanics. The dissipative environment is, however, an issue, but the systems offer a neat platform to explore the role of
 	 decoherence for quantum systems as well.\par 
 	 The unique properties offered by metamaterials and plasmonic structured surfaces can offer new testing grounds for quantum theories. For example, surface plasmons on the interfaces of metals and nonlinear materials can offer deep insights into Anderson localization of interacting particles in two dimensions. Similarly,
 	 nonlinear metamaterials and nonlinear surface plasmonic structures can enable controlled experiments of quantum tunnelling in time retarded systems by studying evanescent electromagnetic waves. The easily modified photonic density modes here offer unique possibilities. The extreme control on the generation of and detection of light (microwaves in particular) make these a good platform for  implementation of many thought experiments that have hitherto been inaccessible in electronic systems.\par 
 In recent times, different types of metamaterials have being studied. Thus, for example, one can have nanowires that use quantum dots as unit cells or artificial atoms arranged as periodic nanostructures. This material has a negative index of refraction and effective magnetism and is simple to build. The radiated wavelength of interest is much larger than the constituent diameter. Photonic Band Gap materials, also known as photonic crystals, are materials which have a band gap due to a periodicity in the materials dielectric properties. A photonic bandgap can be demonstrated with this structure, along with tunability and control as a quantum system. The band gap in photonic crystals represents the forbidden energy range where  photons can not be transmitted through the material. Quantum metamaterial can also be realized using superconducting devices both  with  and without Josephson junctions and are being actively investigated. Recently a superconducting quantum metamaterial prototype based on flux qubits was realized \cite{Macha}.


\section{Optomechanical Systems}

Recent experiments in cavity quantum electrodynamics
(cQED) have explored the interaction of light with atoms
as well as semiconductor nanostructures inside a cavity.
Cavity quantum elctrodynamics is the study of interaction between light confined in a cavity and atoms, where the ambient conditions are such that the quantum nature of light becomes predominant. This field could be traced to the Purcell effect \cite{Purcell}, which is connected to the process of spontaneous emission, a purely quantum effect by which the system transitions from an excited energy state to a lower energy state, emitting, in the process, a quantized amount of energy in the form of a photon. The Purcell effect is the process of enhancement of the systems spontaneous emission rate by its ambient environment. It is dependent on the quality $Q$ factor of the cavity, which is a measure of the ratio of the energy stored to the energy dissipated. A good quality cavity would have a high $Q$ factor.

 Light carries momentum which gives rise to radiation-pressure forces. Recent works have been able to study coupled cavity photons to solid-state mechanical systems containing a
large number of atoms \cite{girvin09}. In these systems, there is an optical cavity, 
with a movable mirror in one end or a micro-mechanical membrane with mechanical effects caused by
light through radiation pressure. Hence, cavity quantum optomechanics has emerged as a very interesting area for revealing quantum features at the mesoscale, where it is 
possible to control the quantum state of mechanical oscillators by
their coupling to the light field. Recent advances in this
area include the realization of quantum-coherent coupling
of a mechanical oscillator with an optical cavity, where
the coupling rate exceeds both the cavity and mechanical motion decoherence rate and laser cooling of a nanomechanical oscillator to its ground state.
Furthermore, new experimental works open up enormous
possibilities in the design of hybrid quantum systems whose
elementary building blocks are physically implemented by
systems of different nature.

\begin{figure*}
	\includegraphics[width=100mm]{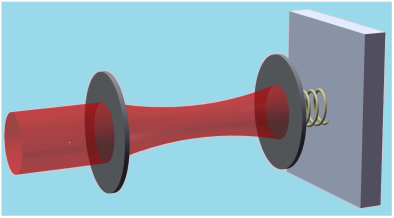}
	\caption{A typical cavity optomechanical system driven by a laser. The left mirror is fixed while  the right mirror is attached to  spring providing the mechanical mode in the system.}
	\label{OptoSystem}
\end{figure*}

Cavity optomechanical systems can provide a natural platform to induce an interaction between mechanical resonators because there is an intrinsic coupling mechanism
between optical and mechanical degrees of freedom.   There has been a lot of interest in the creation of quantum correlations in macroscopic mechanical systems, achieved by means of optomechanical models.  A typical optomechanical system driven by a laser is shown in Fig. \ref{OptoSystem}.

On the one hand, there is the highly sensitive optical detection of small forces,
displacements, masses, and accelerations. On the other hand,
cavity quantum optomechanics promises to manipulate and
detect mechanical motion in the quantum regime using light,
creating nonclassical states of light and mechanical motion.
These tools form the basis for applications in quantum
information processing, where optomechanical devices could
serve as coherent light-matter interfaces, for example, to
interconvert information stored in solid-state qubits into flying
photonic qubits.  At the same time, it
offers a route towards fundamental tests of quantum mechanics
in a hitherto unaccessible parameter regime of size and mass.

\section{PT Symmetry} 
 The concept of Parity-Time (${\rm PT}$) symmetry has played a crucial role in extending  quantum mechanics to the non-Hermitian domain. A non-Hermitian Hamiltonian can also have real eigenvalues if it possess PT symmetry. To be precise, the Hamiltonian $\hat{H}$ with $\hat{H} \ne \hat{H}^\dagger$, where the symbol $\dagger$ stands for hermitian conjugation,  can have a real spectrum of eigenvalues if $(\hat{P} \hat{T})\hat{H} = \hat{H}(\hat{P} \hat{T})$, where $\hat{P}$ and $\hat{T}$ are the Parity and  Time-reversal operators, respectively. They have the following actions on the position ($\hat{x}$) and momentum ($\hat{p}$): $\hat{P} \hat{x}\hat{P}  = - \hat{x}$, $\hat{P} \hat{p} \hat{P}  = - \hat{p}$; $\hat{T} \hat{p} \hat{T} = - \hat{p}$, $\hat{T} \hat{x} \hat{T} =  \hat{x}$ and $\hat{T} \hat{i} \hat{T} = - \hat{i}$. Therefore, for a general Hamiltonian $\hat{H} = \hat{p}^2/2m + \hat{V}(\hat{x}) $ to be PT symmetric, the potential term must satisfy the condition  $ \hat{V}(\hat{x})  = \hat{V}^*(-\hat{x})$.         
 Although the idea of PT symmetry was introduced in quantum mechanics in \cite{bender2007making}, the importance of the phenomenon has been realized recently \cite{Ruter}. Equivalance of a quantum system possesing PT symmetry and to a quantum system having Hermitian Hamiltonian was shown in \cite{mostafazadeh2003exact}. 

A new direction, in the application of PT symmetry, could be the study of non-Hermitian effects in small-scale devices as well as in atomic and molecular systems,
where quantum processes are known to play a significant role. These
include, for instance, driven atomic condensates in cavities, artificial atoms or hybrid quantum systems in cavity quantum electrodynamics as well as coupled optomechanical resonators with gain and loss, where effects such as phonon lasing near exceptional
points could be explored. Initial theoretical studies in this
direction show that the presence of quantum noise leads to significantly different physics as compared to that expected from semiclassical approaches. Novel phases with preserved or {\it weakly} broken PT symmetry appear. Such interactions could in principle also appear in other contexts of quantum optics, such as the prototypical case of an atom interacting with the mode of a light field in an open system.

In \cite{PT-whisprngGalry}, a system was realized whose dynamics is governed by a PT Hamiltonian. Many optomechanical properties have been investigated, such as the cavity optomechanical properties underlying the phonon lasing action,  PT symmetric chaos, cooling of mechanical oscillator, cavity assisted metrology, optomechanically-induced-transparency (OMIT)  and optomechanically induced absorption (OIA). The possibility of the spontaneous generation of photons in PT symmetric systems is illustrated in.

\section{Quasiprobability distributions and Tomography}

\subsection{Quasiprobability distributions}

A very useful concept in the analysis of the dynamics of classical
systems is the notion of phase space. A straightforward extension
of this to the realm of quantum mechanics is however foiled due to
the uncertainty principle. Despite this, it is possible to construct
quasiprobability distributions (QDs) for quantum mechanical systems
in analogy with their classical counterparts.
These QDs are very useful in that they provide a quantum classical
correspondence and facilitate the calculation of quantum mechanical
averages in close analogy to classical phase space averages.
 Nevertheless,
the QDs are not probability distributions as they can take negative
values as well, a feature that could be used for the identification
of quantumness in a system.

The first such QD was developed by Wigner resulting in the epithet
Wigner function ($W$) \citep{wig}.
Another, very well known, QD is the $P$ function whose development
was a precursor to the evolution of the field of quantum optics. This
was originally developed from the possibility of expressing any state
of the radiation field in terms of a diagonal sum over coherent states
\citep{glaub,sudar}. The $P$ function can become singular for quantum
states, a feature that promoted the development of other QDs such
as the $Q$ function  as well as further
highlighted the use of the $W$ function which does not have this
feature. 

A nonclassical state can be used to perform tasks that are classically
impossible. This fact motivated many studies on nonclassical states,
for example, studies on squeezed, antibunched and entangled states.
The interest in nonclassical states has increased with the advent of quantum information processing where several applications of nonclassical states have been reported.
The fields of quantum optics and information have matured to the point
where intense experimental investigations are being made. Both from
the fundamental perspective as well as from the viewpoint of practical
realizations, it is imperative to study the evolution of the system
of interest taking into account the effect of its ambient environment.
This is achieved systematically by using the formalism of Open Quantum
Systems \citep{SBbook}.


\subsection{Tomography}

There is no general prescription for direct
experimental measurement of the quasidistribution functions, such as Wigner function
\cite{wig-diff}.  In general, to detect
the nonclassicality in a system the Wigner function is obtained either
by photon counting or from experimentally measured tomograms \cite{wig-diff}.
Reconstruction of a quantum state from experimentally measured
values is of prime interest for both quantum computation 
and communication. The tomogram is one such candidate
as it is experimentally measurable and is obtained as a probability
distribution.  Further, the quantum
state tomography has its applications in quantum cryptography. 

How to reconstruct a quantum state from experimentally measured
values is of prime interest for both quantum computation 
and communication. Specifically, in \cite{wig-exp} it is strongly established that 
tomography and spectroscopy can be interpreted as dual forms of quantum computation. From the experimental perspective, a quantum state always interacts
with its surroundings. Hence, it is important to consider the evolution of the  tomogram
after taking into account the interaction of the quantum state with
its environment.
\section{Topological photonics}
 Topological phases in condensed matter, that usually arise out of Berry phase effect during adiabatic evolution of states, has taken the centre-stage of research in the condensed matter community for quite some time now.  Interesting findings such as topological insulators, where conducting edge/surface states appear in an otherwise bulk-insulating system or Weyl semimetals, where topologically robust Weyl charges, analogous to magnetic monopoles of the Berry curvature of Bloch bands, appear in pairs within its bulk, are currently being investigated with intense vigour.\par
Over the last decade, there has been a very exciting new development in quantum optics with roots in  condensed matter physics, in particular, topological insulators and quantum Hall effects. This is the field of topological photonics \cite{topolPhotonics, Ozawa},  where externally provided photons in photonic crystals can induce surprising topological effects. As topological insulators are rare among solid-state materials, suitably designed electromagnetic media (metamaterials) can demonstrate a photonic analogue of a topological insulator.  They provide topologically non-trivial photonic states, similar to those that have been identified for condensed-matter topological insulators. The interfaces of these metacrystals support helical edge states, robust against disorders.\par 
Topology is the study of geometrical conserved quantities and its use, in the context of optics, allows the creation of new states of light with interesting properties. Thus, for example, one could have robust unidirectional waveguides allowing light to propagate around defects without back-reflection. This also provides opportunities to realize and exploit topological effects in new ways. The practical implications of topological photonics include the possibility of applications to quantum information processing and topological quantum computing. Another application could be topological lasers, the study of laser oscillation in topological systems.\par
 We have attempted to give a brief overview of the recent developments in the field of quantum optics. Efforts have been made to connect the modern developments with the roots of the subject. The subject has seen an enormous progress in various directions and is believed to provide the testbed for exploring some of the fundamental problems of physics.

\end{document}